\documentclass[aps,prl,twocolumn,superscriptaddress,showpacs]{revtex4}
\usepackage{graphicx}

\begin{document}


\title{Direct observation of hydrodynamic instabilities in driven non-uniform
colloidal dispersions}

\author{Adam Wysocki}

\affiliation{IFF, Forschungzentrum J\"ulich, D-52425 J\"ulich, Germany}
\affiliation{Institut f\"ur Theoretische Physik II, Heinrich-Heine-Universit\"at D\"usseldorf, 
Universit\"atsstrasse 1, D-40225 D\"usseldorf, Germany}

\author{C. Patrick Royall}

\affiliation{School of Chemistry, University of Bristol, Bristol, BS8 1TS, UK }

\author{Roland G. Winkler}
\affiliation{IFF, Forschungzentrum J\"ulich, D-52425 J\"ulich, Germany}

\author{Gerhard Gompper}
\affiliation{IFF, Forschungzentrum J\"ulich, D-52425 J\"ulich, Germany}

\author{Hajime Tanaka}

\affiliation{Institute of Industrial Science, University of Tokyo, 4-6-1 Komaba,
Meguro-ku, Tokyo 153-8505, Japan}

\author{Alfons van Blaaderen}

\affiliation{Soft Condensed Matter Group, Debye Institute for Nanomaterials Science,
Utrecht University, PO Box 80000, 3508 TA Utrecht, The Netherlands}

\author{Hartmut L\"owen}
\affiliation{Institut f\"ur Theoretische Physik II, Heinrich-Heine-Universit\"at D\"usseldorf, 
Universit\"atsstrasse 1, D-40225 D\"usseldorf, Germany}

\email[]{adam@thphy.uni-duesseldorf.de}


\date{\today}

\begin{abstract}

A Rayleigh-Taylor-like instability of a dense colloidal layer under gravity 
in a capillary of microfluidic dimensions is considered. We access  all relevant 
lengthscales with particle-level microscopy and computer simulations which 
incorporate long-range hydrodynamic interactions between the particles. 
By tuning the gravitational driving force, we reveal a mechanism whose growth 
is connected to the fluctuations of specific wavelengths, non-linear pattern 
formation and subsequent diffusion-dominated relaxation. Our linear stability 
theory captures the initial regime and thus predicts mixing conditions, 
with important implications for fields ranging from biology to nanotechnology.

\end{abstract}

\pacs{82.70.Dd, 47.57.ef, 61.20.Ja, 05.40.Jc}

\maketitle

Particulate dispersions have long been subjected to external
fields as a means to separate different constituents; in particular, 
sedimentation is important not only for analytical but also for preparative
purposes \cite{manoharan2003}.
For bulk systems, successful separation depends crucially upon avoiding
hydrodynamic instabilities. The development of microfluidics 
\cite{squires2005} has made it possible to exploit the suppression
of turbulence at small lengthscales in order to design novel separation
devices \cite{huh2007}; on the other hand, this significantly increased 
stability against mechanical perturbations severly limits mixing needed for 
many `lab-on-a-chip' applications. 
Often strong external fields \cite{elmoctar2003}
or complex fabrication \cite{simonnet2005} are required
to produce hydrodynamic instabilities required for efficient mixing. 

Experiments \cite{segre1997,royall2007} and computer simulations 
\cite{padding2008} which study velocity fluctuations have played a crucial 
role in our understanding of how dispersions respond to external driving 
fields, in particular to gravity. The motion of a solute particle
is characterised by a Peclet number $Pe=\tau_{D}/\tau_{S}$, which
is the ratio between the time $\tau_{D}$ it takes a particle to diffuse
its own radius and the time $\tau_{S}$ it takes to sediment the same distance. 
A Peclet number of order unity is the dividing line between colloidal ($Pe\lesssim1$) 
and granular systems ($Pe\gg1$), i.e. $Pe$ measures the importance of Brownian 
motion. All attempts at a quantitative description of sedimentation to date considered a homogenously
distributed dispersion as the initial state. 
For preparative purposes, on the other hand, starting with
a particle-rich layer on top of pure solvent is more relevant as it
enables the separation of particles depending on their sedimentation
coefficient. However, this configuration is unstable with respect to gravity.
The particle velocities become correlated, which leads to emergent
density fluctuations and consequently more rapid sedimentation than 
Stokes' flow alone. It is well known that many practical
particle concentrations develop this Rayleigh-Taylor (RT) like instability.
This provides an avenue by which the system may be successfully
mixed on the one hand, conversely this very mixing, leads, chaotically, 
to a scenario in which separation does not occur. 
For stable separation, it is essential to avoid 
the RT instability. It is possible to use a density gradient to counteract
the instabilities \cite{manoharan2003}. 

The `original' Rayleigh-Taylor instability, which occurs if a heavy, immiscible fluid
layer is placed on top of a lighter one has been intensively studied
for the case of a simple Newtonian fluid both by theory \cite{chandrasekhar},
simulation \cite{kadau2004} and experiment, and 
is observed in granular matter \cite{voeltz2001,vinningland2007_1}, 
in surface-tension dominated colloid-polymer mixtures \cite{aarts2005} 
and in a suspension of dielectric particles exposed to an ac electric field gradient \cite{zhao2008}.

Here we consider a suspension of colloidal hard spheres
(without surface tension) of microfluidic dimensions, in 
which we have access to all relevant length scales, from the single particle level 
to the full system. A systematic study of sedimentation in an 
\emph{inhomogenous} system is presented. We employ three approaches: 
experiment, computer simulation and theory. The experimental realisation is 
provided by confocal microscopy at the single-particle level 
\cite{vanblaaderen1995}, while the simulation is a particle-based mesoscale 
technique \cite{malevanets1999} which captures the direct interactions between 
the colloidal particles, and, crucially, the solvent which mediates the 
hydrodynamic interactions and whose backflow drives the RT instability. 
Our results at short times are modelled with a linear stability 
analysis \cite{chandrasekhar}.

\begin{figure}
\includegraphics[width=8.6cm]{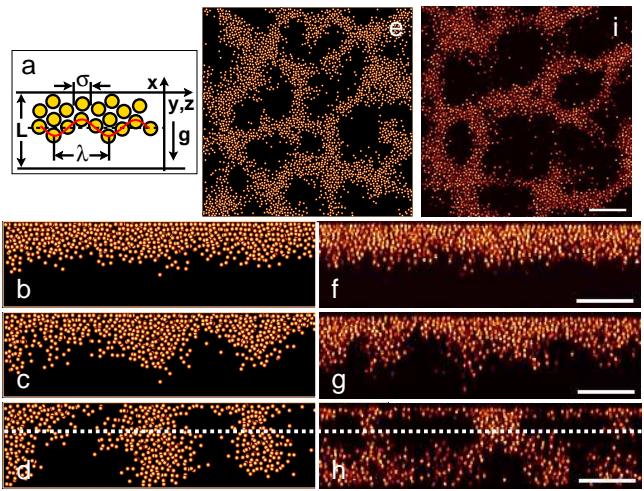}
\caption{\textbf{a}, A schematic illustrating the spatial parameters 
$\sigma$, $\lambda$ and $L$. \textbf{b-e}, Simulation snapshots of a system 
which contains $N=33,858$ colloidal particles and $N_{s}=32,118,397$ solvent particles
(not displayed) in a simulation box with dimensions $L/\sigma=18$
and $L_{y}/\sigma=L_{z}/\sigma=81$. The value of the Peclet number
is $Pe=1.6$. \textbf{b-d}, Time series of the system at time $t/\tau_{S}=3.2$
$(\textbf{b}),$ $6.4$ $(\textbf{c}),$ $9.6$ $(\textbf{d})$. The
snapshots are slices of thickness $2\sigma$ done in the $xy$ plane.
\textbf{e}, Slice of thickness $2\sigma$ in the $yz$ plane at time
$t/\tau_{S}=9.6$. The height of the $yz$ plane is $x/L=2/3$, as
indicated by the dashed line in (\textbf{d}). \textbf{f-i}, Experimental
realisation of the Rayleigh-Taylor-like instability. 
\textbf{f-h}, Time series of images taken with
a confocal microscope in the $xy$ plane for the parameters $\phi=0.15$,
$Pe=1.1$ and $L_{x}/\sigma=18$ at times $t/\tau_{S}=1.43$ $(\textbf{f}),$
$5.5$ $(\textbf{g}),$ $11.22$ $(\textbf{h})$. \textbf{i}, Slice
in the $yz$ plane at a height $x/L=2/3$ (indicated by the dashed
line in (\textbf{h})) at time $t/\tau_{S}=11.22$. In $(\textbf{f-i})$ 
the scale bar denote 40 $\mu$m.}
\label{figSnapShots} 
\end{figure}

The RT instability is thought of as a fluctuation in the interface
between two fluids. Since in a hard-sphere suspension there is no phase 
separation, we consider a continuous density profile, albeit rapidly varying. To capture the
lateral fluctuations, we consider the stability of this density and
associated pressure profile against fluctuations of wavelength $\lambda$
in a horizontal plane perpendicular to gravity. We consider a slit
geometry of height $L$ which is sketched in Fig. \ref{figSnapShots}\textbf{a}.
In the absence of surface tension, the fluctuations of all wavelengths
are in principle unstable, but short wavelength fluctuations are washed
out by diffusion of the colloidal particles and so do not grow exponentially 
\cite{duff1962,kurowski1995}. 

Our linear stability analysis, which reveals the stable and fast-growing wavelengths 
of fluctuations, is based on a continuum hydrodynamics approach where the colloidal 
dispersion is considered as an incompressible one-component fluid with inhomogeneous mass density
$\rho(x)$ and corresponding kinematic viscosity $\nu(x)$ as obtained from the 
Saito representation \cite{ladd1990}. The spatially varying
density profile is given by $\rho(x)=\phi(x)\rho_{c}+(1-\phi(x))\rho_{s}$,
where $\rho_{c}$ and $\rho_{s}$ are the mass densities of the colloidal
particles and the solvent. The colloidal packing fraction profile
$\phi(x)$ is an input from an equilibrated simulation for inverted gravity.
 The stability of the initial density $\rho(x)$
and pressure $p(x)$ profiles against pertubations 
$\delta\rho\propto\delta p\propto\exp\left(i(k_{y}y+k_{z}z)+n(k)t\right)$
with wave number $k=(k_{y}^{2}+k_{z}^{2})^{1/2}$ in the $yz$ plane
and growth rate $n$ is calculated via the linearized Navier-Stokes
equations \cite{chandrasekhar} resulting in the eigenvalue problem  
\begin{eqnarray}
&n\{(\rho u_{x}')'-\rho k^{2}u_{x}\}=\{\nu(u_{x}'''-k^{2}u_{x}')+\nu'(u_{x}''+k^{2}u_{x})\}'\nonumber\\
&-k^{2}\{\frac{g}{n}\rho'u_{x}+2\nu'u_{x}'+\nu(u_{x}''-k^{2}u_{x})\} 
\end{eqnarray}
with the spatial derivative $\dots'=d\dots/dx$, the strength of
the gravitational field $g$ and the fluid velocity field in gravity
direction $u_{x}(x)$. For a system confined between two rigid walls 
we impose $u_{x}=0$ along with the no-slip boundary conditions $du_{x}/dx=0$ at $x=0,L$. 
We account for colloid diffusion by the correction term $n^{*}(k)=n(k)-Dk^{2}$ \cite{duff1962,kurowski1995},
with diffusion constant $D=k_{B}T/3\pi\eta_{s}\sigma$ ($\sigma$ colloid diameter) and dynamic solvent 
viscosity $\eta_{s}$.

In our  computer simulation, which includes solvent-mediated momentum transfer between 
the colloidal particles, we consider a suspension of $N=15,048$ hard sphere particles 
of mass $M$ immersed in a bath of typically $N_{s}=14,274,843$ solvent particles of mass $m$ 
and number density $n_s=N_{s}/V$. The solvent particles are subjected to 
multi-particle collision dynamics \cite{malevanets1999,lamura2001}, which consists 
of two steps. In the streaming step, solvent particles move ballistically 
for time $\delta t$. In the collision step, particles are sorted 
in cubic cells of size $a$, and their velocities relative to the 
center-of-mass velocity of each cell are rotated by an angle $\alpha$ 
around a random axis. We employed the parameters 
$\delta t=0.2\sqrt{ma^2/k_BT}$, $\alpha=3\pi/4$, $n_sa^3=5$ 
and $M=167m$ in order to achieve the hierarchy of time scales 
and the same hydrodynamic numbers as in the experiment, see Ref. \cite{padding2008,ripoll2004} for details.
To enforce no-slip boundary condition on the colloid surface 
and the confining walls a stochastic-reflection method \cite{inoue2002} 
is applied. Statistical averages for time-dependent quantities are performed 
over $200$ independent configurations. 

In our single-particle level confocal microscopy experiments we used polymethylmethacrylate colloids sterically stabilised 
with polyhydroxy-stearic acid. The colloids
were labeled with the fluorescent dye coumarine and had a diameter
$\sigma=2.8$ $\mu$m with around 4\% polydispersity as determined
by static light scattering. To almost match the colloid refractive
index we used a solvent mixture of cis-decalin and cyclohexyl bromide
(CHB), which we tuned to yield different Peclet numbers, owing to
changes in the degree of density mismatch between colloids and solvent.
Specifically, $Pe=1.1$ and $Pe=2.4$ correspond to 80\% and 87.5\% CHB by 
weight respectively. The characteristic time to diffuse a radius 
$\tau_D \approx 29$ s. The data were collected
on a Leica SP5 confocal microscope, fitted with a resonant scanner,
at a typical scan-rate of around 10 s per 3D data set. Prior equilibration
was achieved by placing the suspension overnight such that it sedimented
across a thin (typically $50$ $\mu m$) capillary. The capillary
was then inverted, and the evolution under sedimentation was followed.

\begin{figure}
\includegraphics[width=8.6cm]{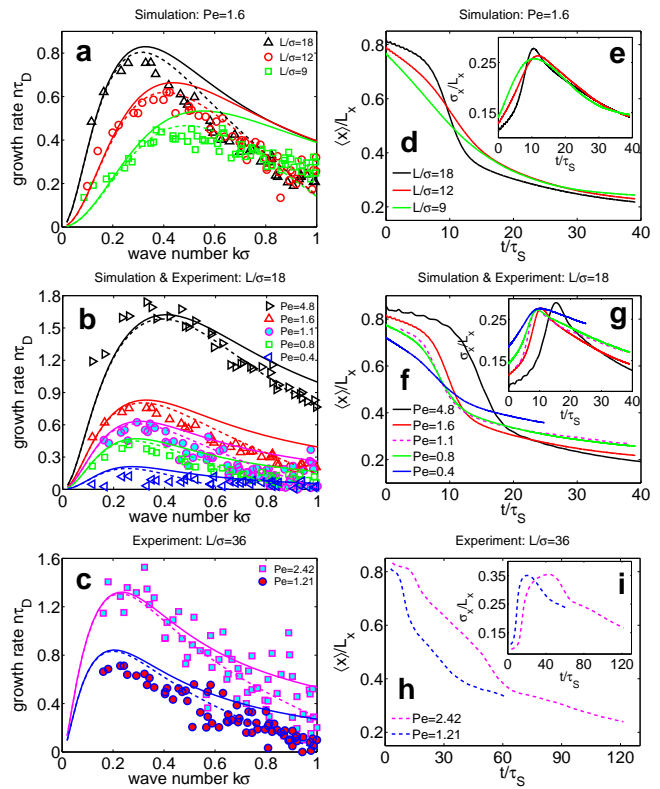}
\caption{\textbf{a-c} Growth rate $n\tau_{D}$ versus wave number 
$k\sigma=2\pi\sigma/\lambda$. \textbf{a},
Simulation results of $n(k)$ for different wall separation distances
$L/\sigma=18,12,9$ and fixed $Pe=1.6$. \textbf{b}, Simulation and
experimental results of $n(k)$ for different Peclet numbers $Pe=4.8,1.6,1.1,0.8,0.4$
and fixed $L/\sigma=18$. \textbf{c}, Experimental results of $n(k)$
for different Peclet numbers $Pe=2.42,1.21$ and fixed $L/\sigma=36$.
The open symbols are the results obtained from simulation, filled
symbols are experimental results, solid lines represent the solutions from the
instabilty analysis and the dashed lines are the same numerical solutions plus 
the diffusion correction. First moment of the colloid density 
$\langle x\rangle/L$ (\textbf{d}),
(\textbf{f}), (\textbf{h}) and second moment of the colloid density
$\sigma_{x}/L$ (\textbf{e}), (\textbf{g}), (\textbf{i}) versus time
$t/\tau_{S}$. Solid lines indicate simulation data whereas the dashed
lines indicate experimental data. \textbf{d,e}, $L/\sigma=18,12,9$
and $Pe=1.6$. \textbf{f,g}, $Pe=4.8,1.6,1.1,0.8,0.4$ and $L/\sigma=18$.
\textbf{h,i}, $Pe=2.42,1.21$ and $L/\sigma=36$.}
\label{figInitialEvolution} 
\end{figure}

We begin our discussion by presenting snapshots of the system, 
in Fig. \ref{figSnapShots}\textbf{b-e} from computer simulation, 
and in Fig. \ref{figSnapShots}\textbf{f-i} from confocal-microscopy. 
The similarity is remarkable, and we note that, at the very
least, our simulation qualitatively reproduces the experiment. 
For a quantitative comparison, we consider the
dispersion relation of wavenumber against growth rate in 
Fig. \ref{figInitialEvolution}\textbf{b}.
The time evolution in the development of the RT instability with
a characteristic wavelength is clear. While snapshots in the gravity
plane (Fig. \ref{figSnapShots}\textbf{b, c, d, f, g,} and
\textbf{h}) illustrate the overall process of sedimentation, snapshots
in the horizontal $yz$ plane show the transient pattern or network-like structure
that results from the RT instability (Fig. \ref{figSnapShots}\textbf{e}
and \textbf{i}). At later times, the network structure decays and
a laterally homogenous density profile develops where the colloids
start to form a layer at the bottom of the cell which becomes more
compact with time. The time evolution is shown in detail in the Movies 1-4, 
see EPAPS Document No. [].

The linear stability analysis predicts the existence of the initially fastest growing wavelengths
in the RT instability. We plot the results of the linear stability
analysis for a range of slit widths $L$ keeping $Pe$ fixed, and
for a variety of Peclet numbers keeping $L$ fixed in Fig. \ref{figInitialEvolution}\textbf{a,}
\textbf{b} and \textbf{c} respectively. The dimensionless growth rates,
$n\tau_{D}$, are plotted as a function of wave number $k\sigma=2\pi\sigma/\lambda$,
where $\lambda$ is the wavelength of the fluctuations as indicated
in Fig. \ref{figSnapShots}\textbf{a}. Without diffusion, fluctuations at all wave numbers 
are unstable, as shown by the solid lines in Fig. \ref{figInitialEvolution}\textbf{a,
b} and \textbf{c}. Due to diffusion, we find that growth
rates at higher wavevectors are suppressed as expected, i.e., diffusion
destroys the Rayleigh-Taylor instability at sufficiently small wavelengths.
We find excellent agreement between the theory with diffusion and
both simulation and experimental data, up to $k\sigma\approx1$,
which is surprising for a coarse-grained continuum description. With
decreasing wall separation $L$, the growth rate $n_{max}$ decreases
and $k_{max}$ increases, see Fig. \ref{figInitialEvolution}\textbf{a}.
Since the fluid velocity in the gravity direction decreases as $e^{-kx}$,
where $x$ is the distance from the interface, only long wavelength
undulations feel the presence of the walls \cite{chandrasekhar,mikaelian1996}.
Figure \ref{figInitialEvolution}\textbf{b} and \textbf{c} show that
for fixed $L$, driving the sedimentation more strongly by increasing
the Peclet number leads to an increase in the wave number
of the fastest-growing undulation $k_{max}$, and the corresponding
growth rate $n_{max}$.

So far we have considered only the linear regime of the instability,
which is valid at small times, when the amplitude of the fluctuations
is smaller than the wavelength. Our experiments and simulations permit
detailed access to all relevant time- and length-scales in the non-linear
regime, where the colloids form foam-like structures in the (confined)
$xz$ plane (Fig. \ref{figSnapShots}\textbf{c},\textbf{g}) 
and a network-like structure in the $yz$ plane (Fig. \ref{figSnapShots}\textbf{e},\textbf{i}) 
appears. Apparently, both continue to exhibit the characteristic length scale $\lambda_{max}=2\pi/k_{max}$
of the fastest growing wavelength in the linear regime.
\begin{figure}
\includegraphics[width=8.6cm]{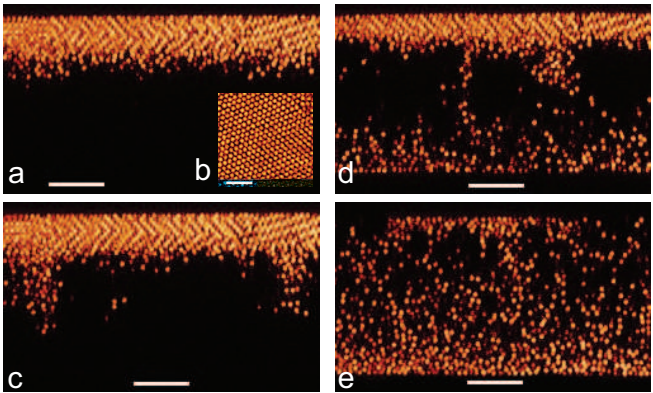}
\caption{Time series of images taken with a confocal microscope in the $xy$ 
plane for the parameters $\phi=0.15$, $Pe=2.42$ and $L_{x}/\sigma=36$ at times 
$t/\tau_{S}=3.1$ $(\textbf{a}),$ $8.1$ $(\textbf{c}),$ $24.6$ $(\textbf{d})$
and $56$ $(\textbf{e})$. The crystalline layers are clearly visible. 
$(\textbf{b})$ is a slice in the $yz$ plane approximately in the middle 
of the colloidal crystal in $(\textbf{a})$. The secondary instability occurs 
at $t/\tau_S\approx20-30$, see $(\textbf{d})$. 
In $(\textbf{a,c-e})$ the scale bar denotes 40 $\mu$m and 20 $\mu$m 
in $(\textbf{b})$.}
\label{figMelting} 
\end{figure}

In order to quantify the different regimes of the instability we use
the first moment of the density $\langle x\rangle$, i.e. the centre
of mass of the colloid coordinates and the second moment of the density
$\sigma_{x}^{2}=\langle x^{2}\rangle-\langle x\rangle^{2}$. Here, $\langle x\rangle$
is a measure of the degree of sedimentation, while $\sigma_{x}$ quantifies
the extent to which the instability spreads out the colloids in the
gravity direction. Three regimes are clearly visible in Fig. \ref{figInitialEvolution}
\textbf{d, f, h, e, g} and \textbf{i}.
Firstly we find the linear regime in which the flat interface develops
undulations and hence $\langle x\rangle$ slowly decreases and $\sigma_{x}$
slowly increases, secondly the non-linear regime where \textbf{`}droplets'
of colloid-rich material fall to the bottom and therefore $\langle x\rangle$
sharply decreases and $\sigma_{x}$ sharply increases, and thirdly
the regime in which the colloids start to form a layer at the bottom
of the cell which becomes more compact with time under settling as
can be seen from the slow decrease of both $\langle x\rangle$ and
$\sigma_{x}$. Clear agreement between simulation and experimental data
can be seen from Fig. \ref{figInitialEvolution} \textbf{f} and \textbf{g}.

In the case of a rather large slit width $L/\sigma=36$, there is
a sufficiently high sediment for a region of colloidal crystal to form, 
see Fig \ref{figMelting} and EPAPS Document No. [] for Movie 5. Since the
crystal has a finite (albeit small) yield stress, the only flow we
observe initially occurs in a thin fluid layer between the crystal and the lower
solvent region via narrow vertical tubes, in marked contrast 
to Fig. \ref{figSnapShots} \textbf{b-d}. 
The crystal melts layer by layer until finally it
becomes sufficiently thin that it peels off the wall in a second instability,
which leads to a change of slope for the $Pe=2.42$ line in 
Fig. \ref{figInitialEvolution}\textbf{h} and \textbf{i} at $t/\tau_{S}\approx20-30$, 
see Fig. \ref{figMelting} \textbf{d}, until most of the particles have sedimented down 
(Fig. \ref{figMelting} \textbf{e}). 
This observation of driven surface melting at the single
particle level has the potential to provide new insight into this
poorly understood phase transition under non-equilibrium conditions.

Using state-of-the-art simulation and experimental techniques, we
have presented a quantitative analysis of a hydrodynamic instability
in a colloidal system at a microfluidic lengthscale. Our
results show excellent agreement between experiment and simulation,
showing that the latter accurately describes the fundamentally
and practically important phenomena caused by hydrodynamic instabilities.
Furthermore, by employing a simple theoretical treatment to the initial
linear behaviour, we find considerable predictive power. The theory can 
flexibly be used to predict conditions for separation and mixing. 
We also note that the theory reveals even the length scale of the
network structure that results from the instability. 
We finally emphasise that the full access and accuracy to all relevant 
length scales in this problem allowed for the observation of novel phenomena, 
not yet explored further, such as the inverse gravity induced crystal melting.

\begin{acknowledgments}
We acknowledge ZIM for computing time. 
A. W. thanks E. W. Laedke and G. Lehmann for help. 
We acknowledge A. A. Louis and J. T. Padding for discussions.
The authors are grateful to Didi Derks for a kind gift 
of PMMA colloids.
A. W., R. G. W., G. G., H. L. and A. v. B. thank the DFG/FOM for support 
in particular via SFB TR6 (projects A3, A4 and D3).
C. P. R. acknowledges the Royal Society for Funding. 
H. T. acknowledges a grant-in-aid from MEXT.
\end{acknowledgments}
\bibliography{rayleigh}

\end{document}